\documentclass[prd,nofootinbib,showkeys]{revtex4}
\usepackage{amsmath,graphicx}

\begin{document}

\title{Weighted finite energy sum rules for the omega meson in nuclear matter}

\author{Birger Steinm\"uller and Stefan Leupold}


\affiliation{Institut f\"ur Theoretische Physik, Universit\"at
Giessen, Germany}

\begin{abstract}
The possible in-medium changes of the properties of an omega meson placed in cold 
nuclear matter are constrained
by QCD sum rules. It is shown that the sum rules cannot fully determine the in-medium
spectral shape of the omega meson. However, for a given parameterization the sum rules
can constrain or correlate the hadronic parameters.
It is shown that weighted finite energy sum rules provide a proper
framework to study directly these constraints/correlations. 
Two typical parameterizations of
possible in-medium omega spectra are analyzed, namely (i) a one-peak structure with
arbitrary position and width of the peak and (ii) a structure with two (narrow) peaks,
caused by the genuine omega meson and a resonance-hole branch. The sum rules provide
for case (i) a mass-width correlation and for case (ii) a correlation between the
peak heights and the peak position of the omega branch. It is also analyzed how the
obtained results depend on the size of the relevant four-quark operator evaluated
with respect to a nucleon. Finally it is argued that a strict vector meson dominance
scenario is not compatible with the sum rules.
\end{abstract}

\maketitle

\section{Introduction}
\label{sec:intro}

The question how hadrons once put in a strongly interacting medium change their 
properties provides a very active field of research. In the language of non-perturbative
QCD, in-medium modifications are indicated by the change of condensates like the
quark condensate \cite{Gerber:1989tt} which provides an order parameter of chiral 
symmetry breaking. On the other hand, changes like the melting of the condensates
do not automatically tell what this means for the properties of a particular hadron 
like its mass or lifetime. The situation is such that the condensates are closer
to QCD as the underlying theory of the strong interaction, whereas the hadron 
properties are closer to experimental observation. The QCD sum rule method is
supposed to bridge that gap by connecting integrals over hadronic spectral functions
with an expansion in terms of quark and gluon condensates. Originally they were
introduced for the vacuum \cite{shif79}, but later on generalized to in-medium situations
\cite{Bochkarev:1986ex}. 

A particularly interesting probe to study in-medium modifications
are neutral vector mesons. The reason is that such mesons can decay into dileptons.
If such a decay happens in the medium the dileptons leave the system untouched by strong
interactions. In that way in-medium information is carried to the detectors. (For an
overview see e.g.~\cite{Rapp:1999ej}.)
The present paper deals with $\omega$-mesons placed in an infinitely extended
system with finite baryonic density. Such a scenario is an idealization of a finite
nucleus or a heavy-ion collision. For simplicity we study $\omega$-mesons which are at
rest with respect to the nuclear medium.

Before we turn to a discussion of in-medium QCD sum rules we present a brief
historical overview concerning the interest in in-medium $\rho$- and $\omega$-mesons: 
On the theoretical side first attempts concentrated on possible mass shifts of
vector mesons \cite{Brown:1991kk,Hatsuda:1992ez} (and the experimental effort was mainly 
directed in heavy-ion collisions).
It was implicitly assumed that e.g.~the widths of the vector mesons are not
drastically changed. Therefore more work was devoted to the $\rho$-meson since a 
long-lived $\omega$-meson would most likely decay outside of the strongly interacting
region. Also, from the experimental point of view it was for a long time impossible to
resolve a narrow peak in the dilepton spectra which would be caused by an $\omega$-meson
with its vacuum width. Recently, however, the interest has shifted to the $\omega$-meson
for the following reasons: On the theory side there appeared calculations which
proposed sizable changes for the in-medium vector meson spectra beyond 
pure mass shifts \cite{herrmann,Chanfray:1993ue,%
Asakawa:1992ht,klingl2,Rapp:1997fs,Peters:1998va,Urban:1998eg,Klingl:1998zj,%
Saito:1998wd,Rapp:1999ej,Post:2000rf,Post:2000qi,Post:2003hu,Bleicher:2000xh,Lutz:2001mi,%
Fuchs:2002vs,Riek:2004kx,Martell:2004gt,MuhlichShklyar:2005}. 
The genuine vector meson peak might get wider due to collisional broadening. In 
addition, even new peaks or bumps might appear caused by collective resonance-hole
excitations. It should be stressed, however, that this issue is far from being
settled. One reason is that a key quantity of such calculations is the forward scattering
amplitude of the respective vector meson with a medium constituent. However, there
is no direct experimental access to such quantities since the vector mesons are not
asymptotic states but resonances. Therefore the calculations are plagued by sizable
model dependences. 
If the life time of vector mesons is significantly reduced
in a medium, the $\omega$-meson might indeed decay inside the medium, while the
$\rho$-meson might get so broad that it can no longer be resolved from the background.
On the experimental side by now also peaks as narrow as the vacuum $\omega$-meson can be
resolved --- at least in dimuon spectra \cite{Damjanovic:2005ni}. 
It is expected that this will also be possible in
presently starting experiments with dielectron spectra \cite{HADES}. 
In addition, heavy-ion
experiments are accompanied by elementary reactions on nuclei. 
Here, the kinematical
situation can be chosen such that the $\omega$-meson is (more or less) at rest with
respect to the nucleus. Hence, even if its life time is long, the $\omega$ would decay
inside the medium. Indeed, a downward shift of strength has been reported for 
$\gamma$-nucleus reactions in \cite{Trnka:2005ey}
for an $\omega$-meson placed in a nucleus. Here the $\omega$ is identified via its
decay into $\pi\gamma$. Note that this decay channel is much suppressed for the $\rho$-meson
(at least in vacuum) so that a clean separation between $\omega$ and $\rho$ seems to be
possible. Dielectron spectra generated in proton-nucleus
collisions are studied in \cite{Ozawa:2000iw,Naruki:2005kd}. 
Also there a downward mass shift of the $\omega$-meson
has been reported. Note, however, that dilepton spectra do not allow for 
a clean separation of (in-medium) $\omega$-mesons from $\rho$-mesons.
In the present work we shall explore
what QCD sum rules tell about the in-medium changes of $\omega$-mesons.

Originally it was expected that the use of nuclear medium QCD sum rules would yield
model independent predictions for in-medium changes of hadronic properties
--- just as the vacuum sum rules yield in an impressive way parameter-free predictions
of vacuum hadronic properties. However, one has to realize that the QCD sum rules
do {\em not} directly yield hadronic properties --- like the mass of a state ---
as a function of the condensates. Unfortunately the connection is rather indirect:
The sum rules connect the condensates with specific integrals 
over the spectral information contained in a correlator of two interpolating quark
currents which carry
the quantum numbers of the hadronic state of interest. {\em Only if} 
this spectral information
is concentrated in a more or less narrow peak the sum rules can predict the peak
position, i.e.~the mass of the hadron under consideration. As we have outlined above 
it is not clear whether the in-medium spectrum of a vector meson is a sufficiently
narrow peak --- it is even not clear whether it has a one-peak structure at all.
In addition, the connection between the hadron and its interpolating quark current
can get more complicated. For vector mesons this concerns the question to which extent
the vector meson dominance assumption still holds in a medium 
(cf.~e.g.~\cite{Friman:1997tc,Post:2000rf}
and also \cite{Harada:2003jx} for possible complications). In general,
for more complicated spectra the sum rules can only yield constraints on these
spectra. For the $\rho$-meson this issue is by now well documented in the literature, 
e.g.~\cite{klingl2,Leupold:1998dg,Leupold:1998bt,Leupold:2001hj,Leupold:2004gh}
For the $\omega$-meson this will become clear below where we will explore 
various typical in-medium parameterizations of the spectral shape.

The present work is not the first one devoted to in-medium QCD sum rules for the 
$\omega$-meson.
Therefore we will comment in the following on the previous approaches. We will discuss
four issues, namely (i) the assumed shape of the spectral information, (ii) the
Landau damping term, (iii) the size of the four-quark condensate and (iv) the type 
of sum rules used. Unfortunately, at least the first three points are intertwined.
We will try our best not to confuse the reader. The motivation for point (i)
has already been given. Point (ii) is important since different works used different
expressions there. To clarify this issue we decided to rederive the proper Landau
damping term in the appendix. Point (iii) concerns the question whether
four-quark condensates factorize into the square of the two-quark condensate.
Already for the vacuum case this constitutes an ongoing debate in the sum rule community.
In addition, the answer to this question might be different for vacuum and in-medium
situations. (For an overview see e.g.~the introduction of \cite{Leupold:2005eq}.) 
The fourth point is somewhat technical. We have decided to postpone the detailed 
discussion to section \ref{sec:cond}.

In the pioneering 
works \cite{Hatsuda:1992ez,Hatsuda:1995dy} a single
narrow peak structure (with the to be determined mass parameter) plus a Landau damping 
contribution was plugged into the sum rules. The result was an in-medium decrease
of the $\omega$-mass. Actually the results agreed for $\rho$- and $\omega$-meson.
The reason was simply that the sum rules were exactly the same and for both mesons
a single-peak structure was assumed --- besides the Landau damping 
contribution. The latter
was correctly derived for the $\rho$-meson. It was assumed that it is the same for the
$\omega$. However, the Landau damping contribution is nine times higher for the $\omega$
than for the $\rho$ as stressed explicitly later in \cite{Dutt-Mazumder:2000ys}
(see also the appendix of the present work). For the $\rho$-meson, the Landau damping
contribution is actually so small that it does not severely influence the result of
a sum rule analysis. A factor nine, however, promotes a negligible contribution
to a dominant one! In principle, the effect of the Landau damping contribution is
an upwards shift of spectral strength. 
Therefore, it was argued in \cite{Dutt-Mazumder:2000ys}
that the $\omega$-mass {\em increases} in nuclear matter. Also in this latter work
a single narrow peak structure was the starting point of the analysis. 
Both in \cite{Hatsuda:1992ez,Hatsuda:1995dy} and
in \cite{Dutt-Mazumder:2000ys} it has been assumed that the in-medium four-quark
condensate more or less factorizes into the square of the in-medium two-quark
condensate. In the present work we will allow for more freedom for the in-medium
spectral information in the $\omega$ channel. One part of our analysis will also cover
the case of a single peak structure, albeit allowing for an arbitrary width. We will
find that the sum rules do not determine both mass and width, but instead provide
a correlation between them. (For the corresponding case of the $\rho$-meson 
see \cite{Leupold:1998dg}.) In a second
part we will study how the sum rules constrain a two-peak structure.
In addition, we will study the sensitivity of our results on the in-medium change
of the four-quark condensate.

Also in \cite{klingl2} the assumption of a narrow peak has been given up. Instead,
a hadronic model for the in-medium spectral function of the $\omega$-meson has
been developed. This model provided a moderate peak broadening and a downward 
mass shift of the $\omega$-meson. It was shown that the calculated spectral function
satisfies the in-medium sum rule. The developed hadronic model has been criticized in
\cite{Friman:1997ce} and therefore modified in \cite{Klingl:1998zj}. Qualitatively,
the findings were still the same: moderate peak broadening and downward mass shift. In
\cite{klingl3} it has been argued that this hadronic model satisfies the sum rules.
However, the Landau damping contribution used there has the correct factor nine,
but an opposite sign as compared to \cite{Dutt-Mazumder:2000ys,klingl2}. Our derivation,
presented in the appendix, agrees in size and sign with \cite{Dutt-Mazumder:2000ys}. 
Indeed, the authors of \cite{klingl3} agree now that the sign of the Landau 
damping term used in \cite{Dutt-Mazumder:2000ys,klingl2}
is the correct one \cite{weise-priv}. 
Nonetheless, a downward mass shift and agreement with the sum rules
has already been reported in \cite{klingl2} where the Landau damping term has the same 
sign as
in \cite{Dutt-Mazumder:2000ys}. Apparently, the findings of \cite{Dutt-Mazumder:2000ys}
(upward mass shift) and of \cite{klingl2} (downward mass shift) seem to contradict
each other. As already speculated in \cite{Dutt-Mazumder:2000ys} this is
probably due to a different treatment of parameters which are intrinsic to the type
of sum rule used. In the present work, where we use a different type of sum rule, 
such ambiguities do not appear as we will discuss below.

The influence of the in-medium behavior of the four-quark condensate has been
studied in detail in 
\cite{Zschocke:2002mp,Zschocke:2003hx,Kampfer:2003sq,Thomas:2005dc}. 
Basically in all previous
works it has been assumed that the in-medium change of the four-quark condensate
is linked to the in-medium change of the two-quark condensate. The
factorization approximation suggests that the drop of the 
four-quark condensate relative to its vacuum value is twice as large as the one for the 
two-quark condensate. 
In \cite{Zschocke:2002mp,Zschocke:2003hx,Kampfer:2003sq,Thomas:2005dc} 
this assumption
has been suspended in favor of an additional parameter which characterizes this drop.
The consequences of the sum rules as a function of this parameter have been studied.
In the present work we will also allow for
an arbitrary in-medium change of the four-quark condensate and study its impact
on our spectral parameterizations. Correlations between hadronic parameters --- on
which we focus in the present work --- 
have not been studied in \cite{Zschocke:2002mp,Zschocke:2003hx,Thomas:2005dc}. 
Mass-width correlations
have been worked out in \cite{Kampfer:2003sq} in the framework of the Borel sum rules.
One purpose of the present work is to establish a new type of sum
rules not used for in-medium studies so far (except for \cite{Leupold:2004gh}), 
namely the weighted finite energy sum rules. Note that for vacuum analyses, the
latter are well established, cf.~e.g.~\cite{maltman,Dominguez:2003dr} and references 
therein.
For in-medium studies, however, usually Borel type sum rules are utilized. We have
decided to postpone the comparison of Borel and other sum rule types to section
\ref{sec:cond} since the important points are easier to discuss with the relevant
formulae at hand. 

The paper is organized in the following way: In the next section we study in-medium
QCD sum rules for the $\omega$-meson channel and in particular their condensate side.
We will start with a Borel sum rule and derive weighted finite energy sum rules
which are directly sensitive to in-medium modifications and
which we will use for the rest of the presented work. Advantages and disadvantages of
different types of sum rules are discussed. In section \ref{sec:had} we study
the constraints provided by the sum rules on typical hadronic parameterizations
of the in-medium spectral information. In particular, we will study
(i) a one-peak structure with
arbitrary position and width of the peak and (ii) a structure with two (narrow) peaks,
caused by the genuine $\omega$-meson and a resonance-hole branch. 
In section \ref{sec:sum} we summarize our results and provide an outlook. The derivation
of the proper Landau damping term is presented in the appendix.

\section{QCD sum rules}
\label{sec:cond}

In this work we study the properties of a vector-isoscalar current 
\begin{equation}
  \label{eq:vecisoscal}
j_\mu := \frac12 \left( \bar u \gamma_\mu u + \bar d \gamma_\mu d \right)
\end{equation}
which is at
rest with respect to the nuclear medium. As outlined e.g.~in \cite{Hatsuda:1993bv}
in-medium QCD sum rules can be obtained from an off-shell dispersion relation
which integrates over the energy at fixed (here vanishing) three-momentum of the current.
We also restrict ourselves to small densities $\rho_N$ 
by using the linear-density approximation. Effectively this means that the current is at
rest with respect to the nucleon on which it scatters.
The Borel sum rule is given 
by \cite{shif79,Hatsuda:1992ez,klingl2,Dutt-Mazumder:2000ys,Zschocke:2002mp}
\begin{eqnarray}
{1\over \pi M^2} \int\limits^\infty_0 \!\! ds \,
{\rm Im} R(s) \, e^{-s/M^2} & = &
{1\over 8\pi^2}\left(1+{\alpha_s\over\pi} \right)
+ {1\over M^4} \, m_q \langle \bar q q\rangle_{\rm med}
+ {1\over 24 M^4} \, \left\langle {\alpha_s \over \pi} G^2 \right\rangle_{\rm med}
+ {1\over 4 M^4} \, m_N a_2 \rho_N 
\nonumber \\
&& 
{}-{56 \over 81 M^6} \, 
\pi\alpha_s \langle {\cal O}^V_4 \rangle_{\rm med}
-{5\over 24 M^6} \, m_N^3 a_4 \rho_N 
+ o(1/M^8) \,. \label{eq:botr}
\end{eqnarray}
Here $\rho_N$ denotes the nuclear density and $M$ the Borel mass. 
The central quantity $R$ is obtained from the current-current correlator
as outlined e.g.~in \cite{Leupold:1998dg}. All other quantities will be specified below.
Note that we have followed the common practice to neglect contributions from non-scalar 
twist-4 operators and from $\alpha_s$ suppressed twist-two operators 
(cf.~e.g.~\cite{Leupold:1998bt} for details). The spectral information ${\rm Im} R$ 
is decomposed into
a hadronic low-energy and a perturbative high-energy part:
\begin{eqnarray}
{\rm Im} R(s) & = &
{9\pi \over 4} \, {\rho_N \over m_N} \, \delta(s)
+ {\rm Im} R_{\rm HAD}(s) \, \Theta(s_0(\rho_N)-s)  
+{1\over 8\pi}\left(1+{\alpha_s\over\pi} \right) \Theta(s-s_0(\rho_N)) 
 \,.
  \label{eq:hacola}
\end{eqnarray}
Here $s_0$ denotes the (density dependent) continuum threshold. We have decomposed 
the low-energy part into the object we want to study, ${\rm Im} R_{\rm HAD}$,
and the Landau damping contribution (first term on the right hand side of
\eqref{eq:hacola}) since the latter can be determined model independently 
(cf.~the appendix). Inserting \eqref{eq:hacola} in \eqref{eq:botr} 
yields \cite{shif79,Hatsuda:1992ez,klingl2,Dutt-Mazumder:2000ys,Zschocke:2002mp}
\begin{eqnarray}
{1\over \pi M^2} \int\limits^{s_0(\rho_N)}_0 \!\! ds \,
{\rm Im} R_{\rm HAD} (s,\rho_N) \, e^{-s/M^2} &=&
{1 \over 8 \pi^2}
\left(1+{\alpha_s\over\pi} \right) 
\left( 1 - e^{-s_0(\rho_N)/M^2} \right) - {9 \over 4 M^2} \, {\rho_N \over m_N} 
\nonumber \\
&& {}+ {1\over M^4} \, m_q \langle \bar q q\rangle_{\rm med}
+ {1\over 24 M^4} \, \left\langle {\alpha_s \over \pi} G^2 \right\rangle_{\rm med}
+ {1\over 4 M^4} \, m_N a_2 \rho_N 
\nonumber \\
  \label{eq:sumrule}
&& {}-{56 \over 81 M^6} \, 
\pi\alpha_s \langle {\cal O}^V_4 \rangle_{\rm med}
-{5\over 24 M^6} \, m_N^3 a_4 \rho_N   \,.
\end{eqnarray}
The four-quark condensate is given by \cite{shif79,Hatsuda:1993bv}
\begin{eqnarray}
\langle {\cal O}^V_4 \rangle &= & 
{81 \over 224} \left\langle 
(\bar u \gamma_\mu \gamma_5 \lambda^a u + \bar d \gamma_\mu \gamma_5 \lambda^a d)^2 
\right\rangle
+ {9 \over 112} \langle 
(\bar u \gamma_\mu \lambda^a u + \bar d \gamma_\mu \lambda^a d)
\sum\limits_{\psi = u, d, s} \bar\psi \gamma^\mu \lambda^a \psi
\rangle  \,.
  \label{eq:fourqdef}
\end{eqnarray}
Using the linear-density approximation we get \cite{Drukarev:1991fs,Hatsuda:1992ez,shif79}
\begin{equation}
  \label{eq:scal2q}
m_q \langle \bar q q \rangle_{\rm med} = m_q \langle \bar q q \rangle_{\rm vac} 
+ m_q \langle N \vert \bar q q \vert N \rangle \rho_N
= -\frac12 F_\pi^2 M_\pi^2 + \frac12 \, \sigma_N \rho_N  \,,
\end{equation}
\begin{equation}
  \label{eq:gluoncond}
\left\langle {\alpha_s \over \pi} G^2 \right\rangle_{\rm med}
= \left\langle {\alpha_s \over \pi} G^2 \right\rangle_{\rm vac}
- {8 \over 11 - {2 \over 3}N_f } m_N^{(0)} \rho_N
\end{equation}
and
\begin{equation}
  \label{eq:scal4q}
\langle {\cal O}^V_4 \rangle_{\rm med} = \langle {\cal O}^V_4 \rangle_{\rm vac} 
+ \langle N \vert {\cal O}^V_4 \vert N \rangle \rho_N
= \langle {\cal O}^V_4 \rangle_{\rm vac} + 2 \kappa \, \langle \bar q q \rangle_{\rm vac}
\, \langle N \vert \bar q q \vert N \rangle \rho_N
= \langle {\cal O}^V_4 \rangle_{\rm vac} 
- \kappa \, \frac{F_\pi^2 M_\pi^2\sigma_N}{2 m_q^2} \,\rho_N  \,.
\end{equation}
Here $\vert N\rangle$ denotes a one-nucleon state (with appropriate normalization). 
We have introduced a quantity $\kappa$ which is basically defined by \eqref{eq:scal4q},
i.e.
\begin{equation}
  \label{eq:defkappa}
\kappa := \frac{\langle N \vert {\cal O}^V_4 \vert N \rangle}
{2 \langle \bar q q \rangle_{\rm vac} \, \langle N \vert \bar q q \vert N \rangle }  \,.
\end{equation}
Our knowledge on the four-quark condensates is rather limited both in vacuum and nuclear
matter (cf.~\cite{Leupold:2005eq} and references therein). 
In the following, we will only need the density dependent part of the four-quark condensate.
In \cite{Leupold:2005eq} it has been shown that in the limit of a large number of quark
colors $N_c$ one gets
\begin{equation}
  \label{eq:kappalnc}
\kappa = 1 + o(1/N_c)  \,.
\end{equation}
For the real world of $N_c=3$ it is not clear how large the deviation of $\kappa$ from unity
actually is. Therefore we will treat $\kappa$ as a free parameter and study the dependence
of our results on $\kappa$. This is close in spirit to \cite{Zschocke:2002mp}.

In principle, we could start our analysis with (\ref{eq:sumrule}). However, there appear
some quantities in (\ref{eq:sumrule}) which have neither to do with the condensates
nor with the low-energy information contained in ${\rm Im} R_{\rm HAD}$: First of all, we
have the Borel mass $M$. As it stands, the sum rule (\ref{eq:sumrule}) does not tell us
for which values of $M$ it is supposed to be valid. Actually we have neglected higher order
condensates $o(1/M^8)$ to obtain (\ref{eq:sumrule}) from \eqref{eq:botr}. 
Therefore, one should not trust
(\ref{eq:sumrule}) for too small values of $M$. On the other hand, if $M$ becomes too large
there is no exponential suppression of larger $s$ on the left hand side of 
(\ref{eq:sumrule}). In other words, the integral becomes more sensitive to the region around
$s_0$. From \eqref{eq:hacola} it is obvious that the modeling of this region is not very
sophisticated. On the other hand, more sophistication would involve more model parameters
which we do not want to have. Therefore, one wants to concentrate on sum rules which are
not very sensitive to the region around $s_0$. For Borel sum rules this is achieved by
not too large values of $M$. These considerations lead to a so-called Borel window.
But the limiting values of this window --- which, in addition, might be density 
dependent --- can only be roughly guessed (cf.~e.g.~\cite{Leupold:1998dg} for details). 
Besides the necessity to choose a proper Borel window we also have the continuum threshold
$s_0$ and its density dependence. It introduces an additional parameter which we are not 
primarily interested in. Finally, the sum rule (\ref{eq:sumrule}) mixes vacuum and
in-medium information while we are only interested in the latter. At least, we want to
make sure that uncertainties in the vacuum description do not influence our conclusions for
the in-medium changes. In the following, we will show that one can get better sum rules
which solve some of the mentioned problems.

To be more sensitive to the in-medium modifications we differentiate the 
Borel sum rule (\ref{eq:sumrule}) with respect to the density:
\begin{eqnarray}
{1 \over \pi M^2} \int\limits_0^{s_0} \!\! ds \, e^{-s/M^2}
\left. {\partial \over \partial \rho_N} {\rm Im}R_{\rm HAD}(s,\rho_N) 
\right\vert_{\rho_N =0} = 
{1 \over M^2} \, c_0 \, e^{-s_0/M^2} + {1\over M^2} \,c_1 + {1\over M^4} \,c_2
+ {1 \over M^6} \, c_3
  \label{eq:bsr}
\end{eqnarray}
with
\begin{subequations}
\label{eq:defc2c3}
\begin{eqnarray}
  \label{eq:defc0}
  c_0 & = & \left[{1 \over 8 \pi^2}
\left(1+{\alpha_s\over\pi} \right) - \frac1\pi {\rm Im}R_{\rm HAD}(s_0,\rho_N=0)\right] 
\, s_0'
\,,
\\[0.5em]
  \label{eq:defc1}
  c_1 & = & -\frac{9}{4 m_N} \,,
\\[0.5em]
  \label{eq:defc2}
  c_2 & = & {m_N a_2 \over 4} + {\sigma_N \over 2} - \frac{m_N^{(0)}}{27} \,,
\\[0.5em]
  \label{eq:defc3}
  c_3 & = & \kappa \, \frac{28 \pi \alpha_s F_\pi^2 M_\pi^2 \sigma_N}{81 \, m_q^2}
  - {5 \over 24} \, m_N^3 a_4  \,,
\end{eqnarray}
\end{subequations}
\begin{equation}
  \label{eq:defs0vac}
  s_0 = s_0(\rho_N = 0)
\end{equation}
and
\begin{equation}
  \label{eq:s0der}
s_0'= \left. {d s_0 \over d \rho_N} \right\vert_{\rho_N =0}  \,.
\end{equation}
Next we rewrite (\ref{eq:bsr}):
\begin{eqnarray}
{1 \over \pi} \int\limits_0^{s_0} \!\! ds \, e^{(s_0-s)/M^2}
\left. {\partial \over \partial \rho_N} {\rm Im}R_{\rm HAD}(s,\rho_N)
\right\vert_{\rho_N =0}
= c_0  + c_1 \, e^{s_0/M^2} + {1\over M^2} \,c_2 \, e^{s_0/M^2} 
+ {1 \over M^4} \, c_3 \, e^{s_0/M^2} \,,
  \label{eq:bsr2}
\end{eqnarray}
expand both sides in powers of $1/M^2$ and compare the corresponding coefficients on
right and left hand side:
\begin{subequations}
\label{eq:FESR}
  \begin{eqnarray}
    \label{eq:wFESR0}
{1 \over \pi} \int\limits_0^{s_0} \!\! ds \, 
\left. {\partial \over \partial \rho_N} {\rm Im}R_{\rm HAD}(s,\rho_N)
\right\vert_{\rho_N =0} 
& = & c_0 + c_1  \,,
\\
    \label{eq:wFESR1}
{1 \over \pi} \int\limits_0^{s_0} \!\! ds \, (s_0-s)
\left. {\partial \over \partial \rho_N} {\rm Im}R_{\rm HAD}(s,\rho_N)
\right\vert_{\rho_N =0} 
& = & c_1 s_0 + c_2  \,,
\\
    \label{eq:wFESR2}
{1 \over \pi} \int\limits_0^{s_0} \!\! ds \, (s_0-s)^2 
\left. {\partial \over \partial \rho_N} {\rm Im}R_{\rm HAD}(s,\rho_N)
\right\vert_{\rho_N =0} 
& = & c_1 \, s_0^2 + 2 c_2 \, s_0 + 2 c_3  \,.
  \end{eqnarray}
\end{subequations}
In this way we have obtained weighted finite energy sum rules. The advantage of
finite energy type sum rules as compared to Borel sum rules lies in the fact that
with the former we have got rid of the Borel mass and the problem how to determine
a reliable Borel window etc. (cf.~our discussion above). 
On the other hand, the standard finite energy sum rules (as applied to the in-medium case
e.g.~in \cite{klingl3}) are rather sensitive to
the modeling of the transition region from the hadronic part ${\rm Im}R_{\rm HAD}$
to the continuum (see also e.g.~\cite{Leupold:2001hj} and references therein). Indeed,
the first equation (\ref{eq:wFESR0}) is plagued by that problem. The latter two
equations, however, are not since the transition region is suppressed by powers of
$(s_0-s)$ \cite{maltman}. Therefore (\ref{eq:wFESR1}) and (\ref{eq:wFESR2}) are
more reliable as they are insensitive to details of the
threshold modeling at $s_0$. Hence these weighted finite energy sum rules combine the
advantages of Borel and standard finite energy sum rules. In general, the disadvantage 
is that there are only two properly weighted finite energy sum rules as compared to 
three standard finite energy sum rules. In our case, however, this does not reduce
the available information: The in-medium change of the threshold parameter encoded in
$s_0'$ is anyway unknown {\it a priori}. Fortunately, it only appears in the first 
(anyway less reliable) sum rule (\ref{eq:wFESR0}). The two preferable sum rules
(\ref{eq:wFESR1}) and (\ref{eq:wFESR2}) are independent of $s_0'$. 
We shall use them for the subsequent analysis. 
Note that the {\em vacuum} threshold $s_0$
appears in (\ref{eq:wFESR1}) and (\ref{eq:wFESR2}). This, however, can be fixed by
an independent vacuum sum rule analysis which is free of all in-medium uncertainties.
For the actual calculation we adopt the point of view of 
\cite{klingl3,Marco:1999xz,Leupold:2003zb} and use 
$s_0 \approx (4 \pi f_\pi)^2 \approx 1.3\,$GeV$^2$ with the pion decay constant
$f_\pi$. Note that one obtains numerically the same value if one takes the arithmetic
average of the squared masses of the $\omega$ and of its first excitation.
We will study the sensitivity of our results with respect to $s_0$ below.

We would like to stress again that the sum rules (\ref{eq:wFESR1}) and (\ref{eq:wFESR2}) 
constitute a big step forward in the sum rule analysis of in-medium properties:
First, we are directly sensitive to in-medium changes in contrast to
traditional analyses \cite{Hatsuda:1992ez,klingl2,Zschocke:2002mp}
which study vacuum plus medium contributions. Second, we have got 
rid of all problems how to properly define a reasonable Borel window
(cf.~also the discussion in \cite{Hatsuda:1995dy,Mallik:2001gv}). Third, we still share
with Borel type sum rules the feature that we are less sensitive to the modeling of the
continuum threshold.

\section{Hadronic parameterizations}
\label{sec:had}

We now turn to the left hand side of the sum rules (\ref{eq:sumrule}) or 
(\ref{eq:FESR}). It is
well-known that the vector-isoscalar current $j_\mu$ strongly couples to the 
$\omega$-meson. In the vector meson dominance (VMD) picture which is phenomenologically
rather successful it is even assumed that
all the interaction of $j_\mu$ with hadrons is mediated by the $\omega$-meson 
\cite{sakuraiVMD}. 
In the following we will rather study simple parameterizations for the current-current
correlator and not detailed hadronic models. Of course, the omega meson will play a 
prominent role here. We will come back to the issue of VMD below.

For all numerical evaluations we will use the values given in table \ref{tab:tabquant}.
All plots are for normal nuclear matter density $\rho_N = \rho_0 \approx 0.17\,$fm$^{-3}$
except where otherwise stated.
\begin{table}[htbp]
  \centering
  \begin{tabular}{|c|c|c|}
\hline
quantity & size & ref. \\ \hline \hline
$m_N $ & $940 \,$MeV & \cite{pdg04} \\ \hline
$m_N^{(0)} $ & $750 \,$MeV & \cite{klingl2} \\ \hline
$a_2$ & 0.9 & \cite{Hatsuda:1992ez} \\ \hline
$a_4$ & 0.12 & \cite{Hatsuda:1992ez} \\ \hline
$\sigma_N$ & $45 \,$MeV & \cite{Hatsuda:1992ez} \\ \hline
$m_q$ & $6 \,$MeV & \cite{pdg04} \\ \hline
$\alpha_s$ & 0.36 & \cite{Hatsuda:1992ez} \\ \hline
$F_\pi$ & $92.4 \,$MeV & \cite{pdg04} \\ \hline
$M_\pi$ & $140  \,$MeV & \cite{pdg04} \\ \hline
$\kappa$ & 0 \ldots 10 & -- \\ \hline
$s_0$ & $1.33\,$GeV$^2 \pm 10\%$ & --  \\ \hline
$M_\omega$ & $783 \,$MeV & \cite{pdg04} \\ \hline
$M_{rh}$ & $590 \,$MeV & \cite{pdg04} \\ \hline
  \end{tabular}
  \caption{Sizes of all relevant quantities.}
  \label{tab:tabquant}
\end{table}

\subsection{Warm up: single narrow peak}
\label{sec:snp}

\begin{figure}[htbp]
  \centering
    \includegraphics[keepaspectratio,width=0.49\textwidth]{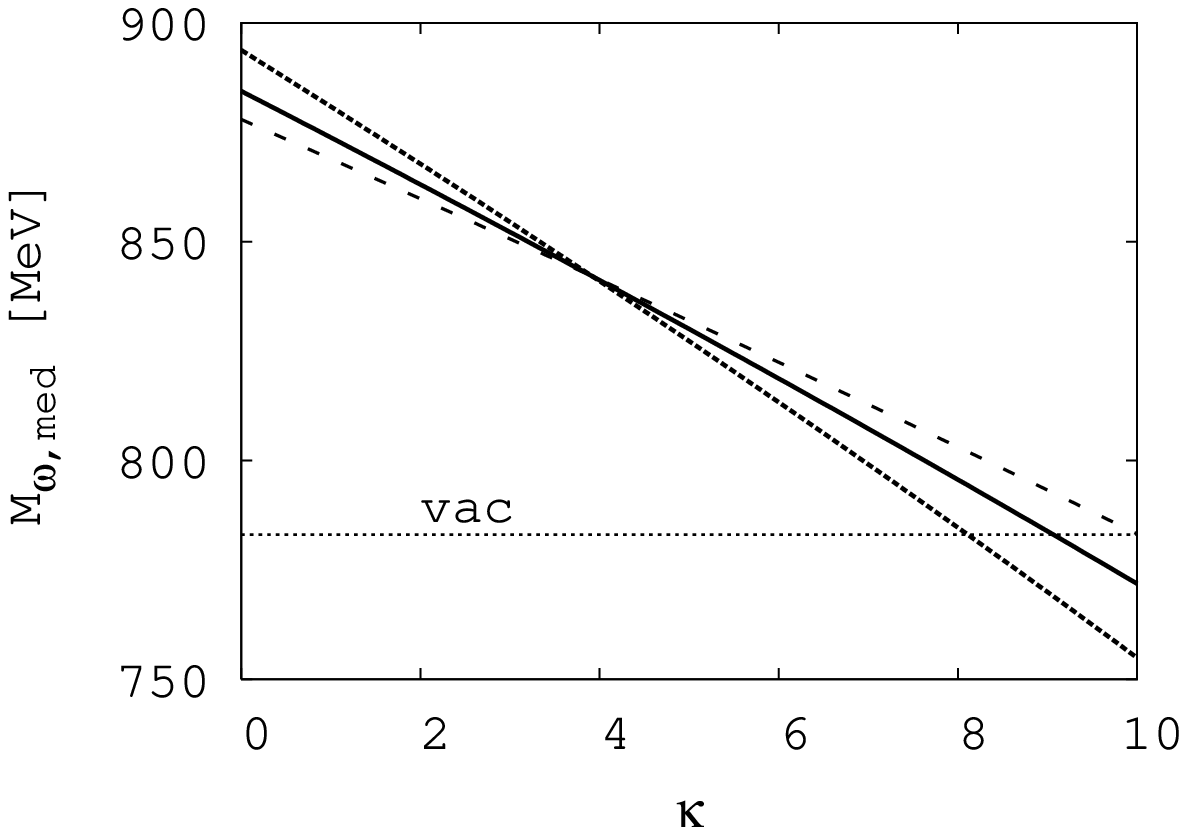}
    \includegraphics[keepaspectratio,width=0.49\textwidth]{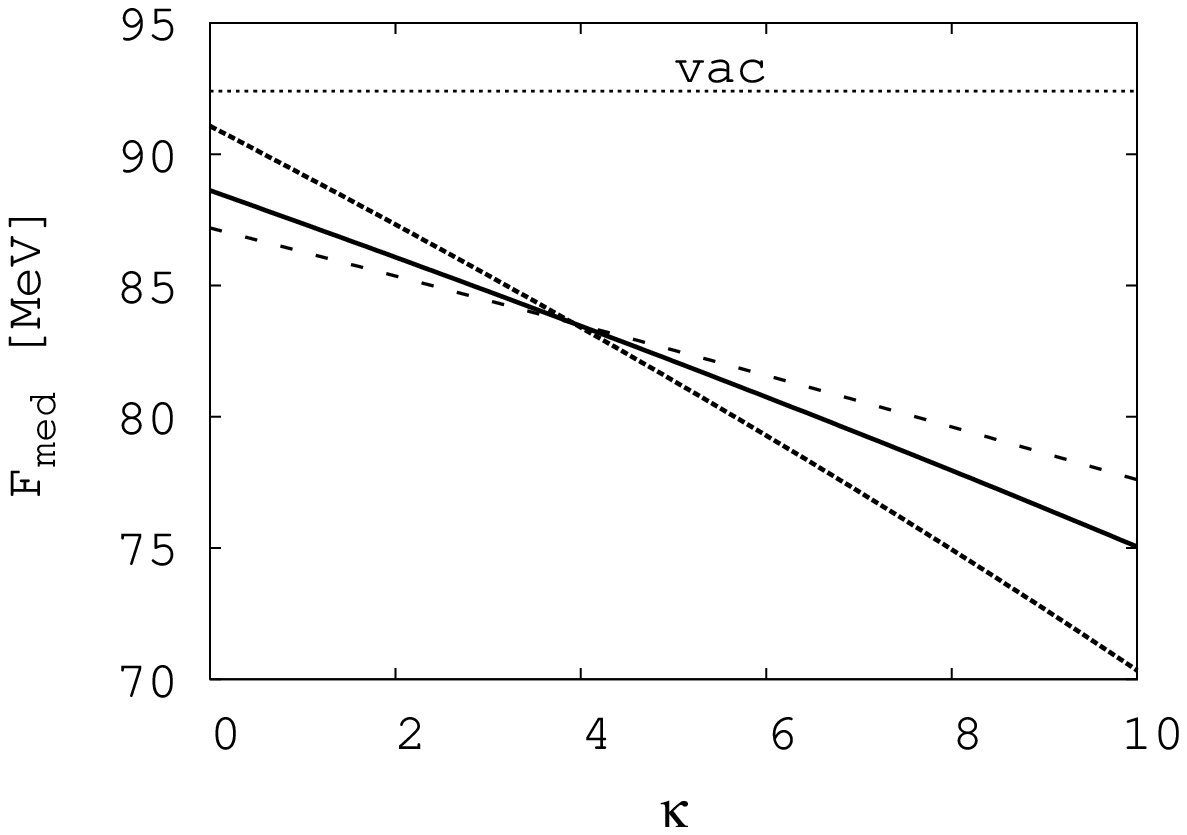}
  \caption{In-medium mass \eqref{eq:defmass} (left panel) and 
in-medium strength \eqref{eq:defstr} (right panel)
of the $\omega$-meson for normal nuclear matter density 
$\rho_N = \rho_0 \approx 0.17\,$fm$^{-3}$ as functions of $\kappa$ for different values
of the vacuum continuum threshold $s_0 = 1.33\,$GeV$^2$ (full lines), 
$1.20\,$GeV$^2$ (dotted), $1.46\,$GeV$^2$ (dashed). The respective vacuum value is denoted
by the horizontal line. Note that the input is a single narrow 
peak structure \eqref{eq:imrhad-onp}.}
  \label{fig:narpeak}
\end{figure}

To make contact with previous works we explore as a first parameterization a spectral
structure with a single narrow peak: 
\begin{equation}
  \label{eq:imrhad-onp}
{\rm Im}R_{\rm HAD}(s,\rho_N) = 
2\pi (F_\pi^2 + c_F \rho_N) \, \delta(s-M_\omega^2-c_M \rho_N)  \,.
\end{equation}
Our choice for the spectral strength in vacuum is motivated in \cite{Leupold:2005ep}.
It yields a good description of the cross section $e^+ e^- \to \omega$. Obviously,
from the two weighted finite energy sum rules (\ref{eq:wFESR1}) and (\ref{eq:wFESR2})
one can in principle determine two quantities, which are in the present case 
$c_F$ and $c_M$. Of course, the results depend on the input parameters given in
table \ref{tab:tabquant}, especially on $s_0$ and $\kappa$. Instead of $c_F$ and $c_M$
we introduce the following physically more intuitive parameters: the
in-medium mass
\begin{equation}
  \label{eq:defmass}
M_{\omega,\rm med} = \sqrt{M_\omega^2+c_M \rho_N}
\end{equation}
and the in-medium strength
\begin{equation}
  \label{eq:defstr}
F_{\rm med} = \sqrt{F_\pi^2 + c_F \rho_N}  \,.
\end{equation}
These quantities are depicted in figure \ref{fig:narpeak} as functions of $\kappa$.

We observe that the mass is increased except for very high values of $\kappa$. This 
finding is in qualitative agreement 
with \cite{Dutt-Mazumder:2000ys,Zschocke:2002mp,Thomas:2005dc}.
Quantitatively, there are some differences which are due to the use of different types
of sum rules --- Borel sum rules 
in \cite{Dutt-Mazumder:2000ys,Zschocke:2002mp,Thomas:2005dc},
weighted finite energy sum rules here. As we have already stressed we prefer the use of 
the latter. It is well known that the drop of the four-quark condensate leads to
a lowering of the mass \cite{Hatsuda:1992ez}, while the Landau damping term leads
to an increase \cite{Dutt-Mazumder:2000ys}. The competition between these two effects 
is responsible for a rising mass at low values of $\kappa$ and a 
decreasing mass for large $\kappa$. We also want to stress that an increase of $\kappa$
by an order of magnitude leads to a change in the mass of only about 10\%. In other words,
with a rough idea about the four-quark condensate one gets a quite good prediction for
the mass. On the other hand, it is complicated to deduce from a measured mass a precise
value for $\kappa$. 

Concerning the input parameter $s_0$ we observe that our results in general do depend
on it, albeit this dependence is not very drastic. As a curiosity we note that there
is one particular value of $\kappa \approx 4$ where there is no $s_0$-dependence. We
do not think that there is any deeper meaning to this point.

Note that we do not study the sensitivity of our results on the other input parameters
of table \ref{tab:tabquant}
for the following reason: The role of $\sigma_N$ and $m_N^{(0)}$ in \eqref{eq:defc2} is
subleading as compared to the $a_2$ term. They even cancel to a large extent. 
The value of $a_2$,
on the other hand, is well determined from deep 
inelastic scattering \cite{Hatsuda:1992ez}. In \eqref{eq:defc3} the quantities 
$m_q$, $\alpha_s$ and $\sigma_N$ are accompanied by 
$\kappa$ which is varied in a large range anyway. Hence, all uncertainties in these
quantities are effectively studied by our variation of $\kappa$. 

In contrast to the mass the in-medium strength is always smaller than its vacuum value
for all values of $s_0$ and $\kappa$ studied here. The strength of the peak is actually
important for the issue of VMD. This, however, is better discussed in a more general
framework where one allows for a width of the peak. We now turn to such a more general
parameterization.

\subsection{Single peak with width}
\label{sec:swp}

Instead of the $\delta$-function present in (\ref{eq:imrhad-onp}) we now allow for a
broad spectral information. Details of the formalism can be found 
in \cite{Leupold:1998dg} for the corresponding case of the $\rho$-meson. The spectral
information is now assumed to be
\begin{equation}
  \label{eq:imrhad-owp}
{\rm Im}R_{\rm HAD}(s,\rho_N) = 
\pi V(\rho_N) \, {{\cal A}(s,\rho_N) \over s}
\end{equation}
with
\begin{equation}
  \label{eq:spec}
{\cal A}(s,\rho_N) = {1 \over \pi} {\sqrt s \,\Gamma_{\rm med}(s) \over 
(s-M_\omega^2-c_M \rho_N)^2+s\,(\Gamma_{\rm med}(s))^2} \,.
\end{equation}
Note that for simplicity we neglect a possible $s$-dependence of the mass parameter. 
We also neglect the (small) vacuum width of the $\omega$-meson. 
\begin{figure}[htbp]
  \centering
    \includegraphics[keepaspectratio,width=0.49\textwidth]{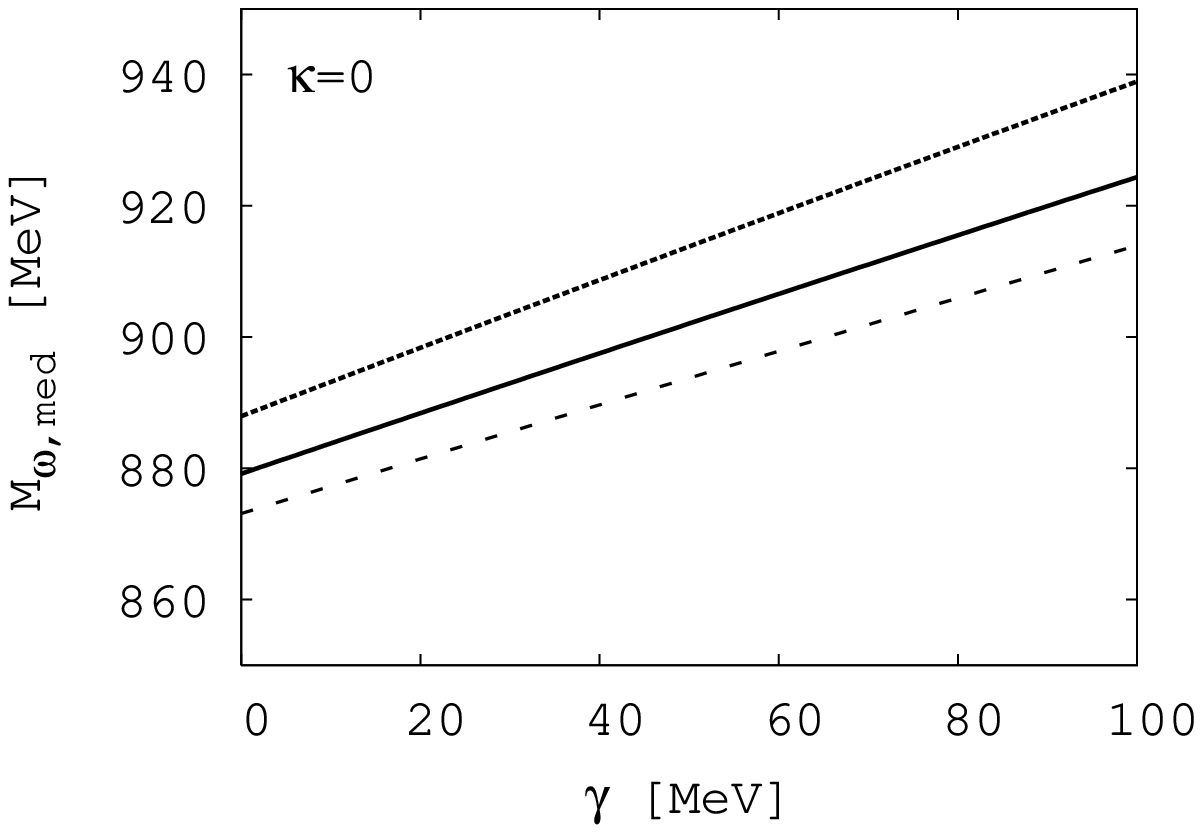}
    \includegraphics[keepaspectratio,width=0.49\textwidth]{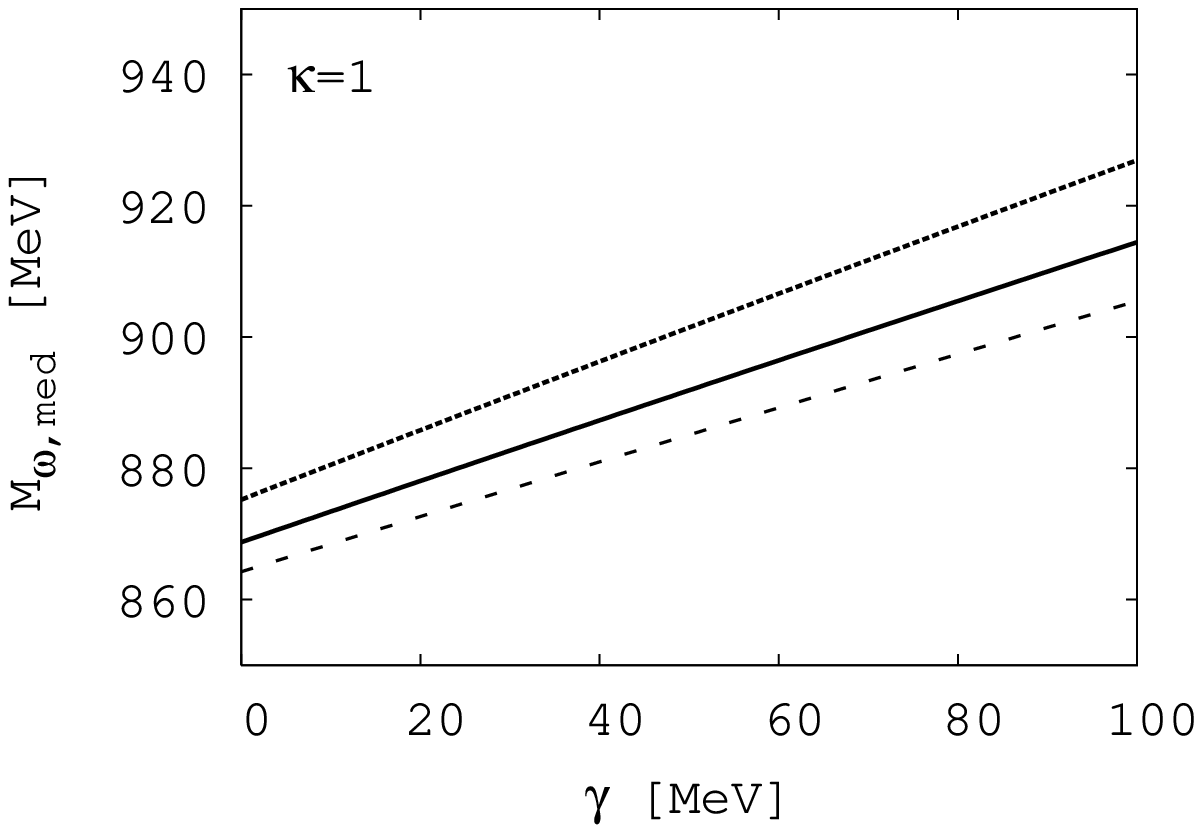}
    \includegraphics[keepaspectratio,width=0.49\textwidth]{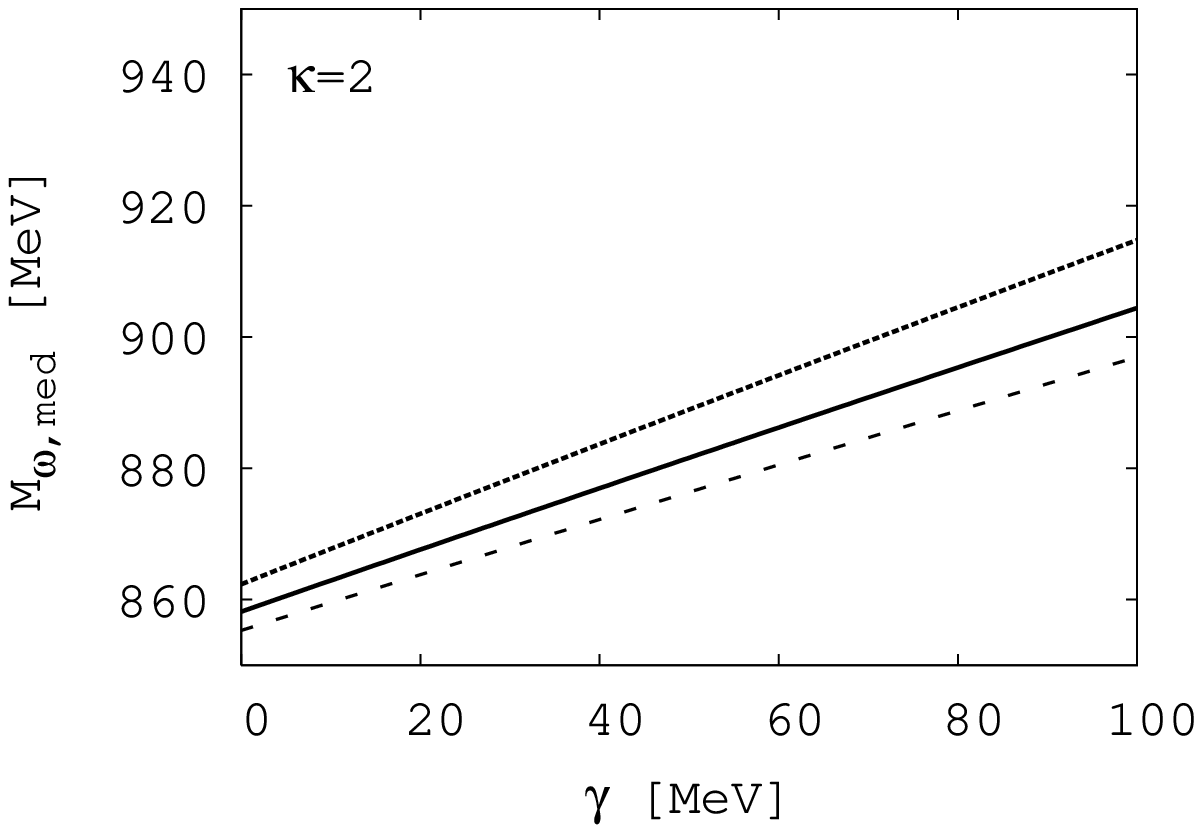}
  \caption{Correlation between in-medium mass \eqref{eq:defmass} and in-medium width 
$\gamma$ of the $\omega$-meson for normal nuclear matter density 
$\rho_N = \rho_0 \approx 0.17\,$fm$^{-3}$ for different values
of the vacuum continuum threshold $s_0 = 1.33\,$GeV$^2$ (full lines), 
$1.20\,$GeV$^2$ (dotted), $1.46\,$GeV$^2$ (dashed) and for different values of the
medium dependence of the four-quark condensate $\kappa =0$ (top, left panel),
1 (top, right), 2 (bottom). 
Note that the input is a single peak structure with width \eqref{eq:imrhad-owp}.}
  \label{fig:broadpeak}
\end{figure}

If the width $\Gamma_{\rm med}$ (which will be further specified below) becomes small
we have to recover the previous case discussed in subsection \ref{sec:snp}. This enables
us to fix the vacuum value of $V(\rho_N)$ in (\ref{eq:imrhad-owp}): Indeed,
${\cal A}$ becomes a $\delta$-function for vanishing $\Gamma_{\rm med}$. 
Comparing then \eqref{eq:imrhad-onp} and \eqref{eq:imrhad-owp}
for the vacuum case one gets
\begin{equation}
  \label{eq:FFpi}
V_0 := V(\rho_N=0) = 2 F_\pi^2 M_\omega^2  \,.
\end{equation}
Having determined the vacuum value, we assume (in line with the linear-density
approximation) that $V(\rho_N)$ scales linearly with $\rho_N$. This leaves us with one
free parameter to characterize $V(\rho_N)$.

Since the integrals appearing in (\ref{eq:FESR}) are obviously sensitive to
the behavior of ${\cal A}$ for small values of $s$ it is important to model the 
threshold behavior of $\Gamma_{\rm med}$ in a physically reasonable way. 
For finite nuclear density the scattering
with nucleons influences the spectral function. 
If the Fermi motion of the nucleons is neglected
the threshold for the spectral function of an $\omega$-meson {\it at rest} is 
given by the mass of one pion since the lightest pair of particles which can
be formed in an $\omega$-nucleon collision is a nucleon and a pion. The threshold
behavior is dominated by the lowest possible partial wave. Without any additional 
constraint from the intermediate state formed in the $\omega$-nucleon collision
we assume it to be an s-wave state. Hence we get
\begin{equation}
  \label{eq:thresmed}
  \Gamma_{\rm med}(s) \sim (s-M_\pi^2)^{1/2} \,\Theta(s-M_\pi^2) 
\end{equation}
and thus 
\begin{equation}
  \label{eq:gammamed}
  \Gamma_{\rm med}(s) = \gamma \,
\left( {1- {\displaystyle M_\pi^2 \over \displaystyle s} \over 
1- {\displaystyle M_\pi^2 \over \displaystyle M_\omega^2} } \right)^{1/2}
 \,\Theta(s-M_\pi^2)
\end{equation}
with the free parameter $\gamma$, the on-shell width. We stress again that
we have neglected the small vacuum width of the $\omega$ in (\ref{eq:spec}). 
Consequently we take $\gamma \sim \rho_N$. 
Note that strictly
speaking the in-medium on-shell width should be taken with respect to the in-medium
mass instead of the vacuum mass in (\ref{eq:gammamed}). 
Therefore, it might be a misnomer to call $\gamma$ the
on-shell width. However, since $\gamma$ is already linear in the density, replacing
the vacuum by the in-medium mass in the width formula (\ref{eq:gammamed})
is beyond the accuracy we are working.

In total, our parameterization (\ref{eq:imrhad-owp}) contains three free parameters:
the in-medium change of the mass (\ref{eq:defmass}), the in-medium change of $V(\rho_N)$
and the width $\gamma$. 
Obviously, our two sum rules (\ref{eq:wFESR1}) and (\ref{eq:wFESR2}) can hardly determine
all three parameters. What we can get, however, are correlations between these
parameters. E.g.~for given width the other two parameters can be determined. 
Most interesting is the mass-width correlation which we show in 
figure \ref{fig:broadpeak}. Corresponding correlations, e.g.~for strength and width
could also be obtained, but are not displayed explicitly. We come back to a discussion
of the strength below.
\begin{figure}[htbp]
  \centering
    \includegraphics[keepaspectratio,width=0.49\textwidth]{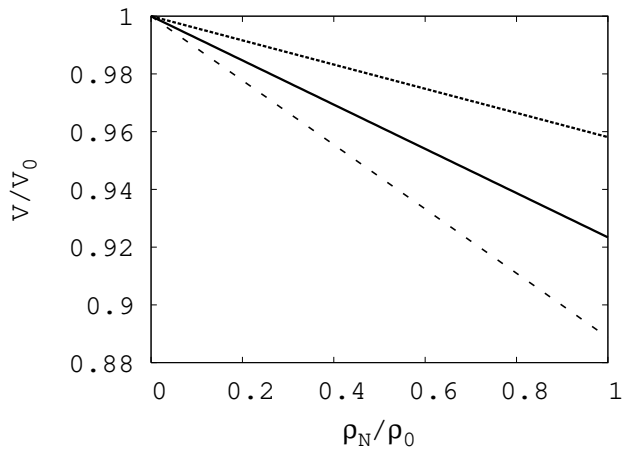}
  \caption{Strength $V$ of \eqref{eq:imrhad-owp} normalized to its vacuum value $V_0$
\eqref{eq:FFpi} as a function of the density $\rho_N$ (normalized to 
normal nuclear matter density $\rho_0 \approx 0.17\,$fm$^{-3}$) for different values
of the width $\gamma(\rho_0) = 0$ MeV (dashed line), 30 MeV (full), 60 MeV (dotted).
We used $s_0 = 1.33\,$GeV$^2$ and $\kappa = 1$.}
  \label{fig:novmd}
\end{figure}

From figure \ref{fig:broadpeak} we observe first of all that the results depend to 
some extent on the vacuum continuum threshold $s_0$ and on the parameter $\kappa$ 
characterizing the in-medium change of the four-quark condensate. Especially, for
increasing $\kappa$ the masses move down. This is exactly in line with the previous
finding. Indeed, the respective mass for
$\gamma =0$ agrees with the corresponding one determined in the previous subsection, 
cf.~figure \ref{fig:narpeak}. When the width increases, the in-medium mass is pushed
up. Qualitatively, this is easy to explain: For the values of $\kappa$ displayed in
figure \ref{fig:broadpeak} the sum rules prefer masses higher than the vacuum one,
as seen in figure \ref{fig:narpeak}. More generally, one might say that the sum rules
prefer accumulation of spectral strength for invariant masses $\sqrt{s}$ 
above the vacuum mass. Naively, one might think that the introduction of a width
does not change the center of the mass accumulation. This, however, is not true:
Due to the weighting factors $(s_0-s)^n$ in the sum rules the contributions of lower
invariant masses are weighted sizably more than the ones from higher invariant masses.
Therefore, the introduction of a width effectively shuffles strength to lower invariant
masses. To account for that, i.e.~to satisfy the same sum rules for the zero-width
and for the finite-width case, the center of the spectral distribution has to shift
upwards. We see exactly this effect in figure \ref{fig:broadpeak}. Similar findings
have been obtained in \cite{Kampfer:2003sq}. Again, quantitative differences are
due to the use of Borel sum rules there and weighted finite energy sum rules
here.

Next we turn to a discussion of the strength $V(\rho_N)$ given in (\ref{eq:imrhad-owp}).
It is important to stress that in a pure VMD scenario all hadronic effects on the 
current (\ref{eq:vecisoscal}) are mediated by the $\omega$-meson. Therefore, all 
in-medium modifications concern the $\omega$-meson, i.e.~the spectral function in
(\ref{eq:imrhad-owp}). In other words, in a pure VMD scenario
$V(\rho_N)$ would be constant, i.e.~independent of the
density. In an extended scenario $V$ might not only depend on the density, but also 
on the invariant mass (squared) $s$ (cf.~e.g.~\cite{Friman:1997tc}). 
Such an additional dependence --- where the details are also 
model dependent --- has been disregarded for simplicity in \eqref{eq:imrhad-owp}.
Figure \ref{fig:novmd} shows curves of $V(\rho_N)$ for different values of the width 
$\gamma$
chosen such that $\gamma(\rho_0) = 0, 30, 60$ MeV. Curves for different values of
$s_0$ and $\kappa$ look qualitatively similar. Obviously a strict VMD scenario is 
excluded by the sum rules. This is an interesting aspect to which not much
attention has been paid in the past --- at least not in the sum rule context. Concerning
hadronic models we refer to \cite{Friman:1997tc,Post:2000rf} for a discussion 
of this issue.
An in-medium fate of VMD has also been found in the framework of hidden local symmetry
\cite{Harada:2003jx}.

\subsection{Two narrow peaks}
\label{sec:tnp}

The purpose of the present subsection is to stress that a one-peak structure might
not be the only possibility for an in-medium spectral information. Indeed, as has been
worked out e.g.~in \cite{Lutz:2001mi,Fuchs:2002vs,MuhlichShklyar:2005} 
at least two peaks 
are conceivable in the spectral function of the $\omega$-meson and therefore also
in the spectral distribution of the current-current correlator discussed here. The
additional peak is caused by the excitation of a nucleon hole and a baryonic resonance.
Resonances in the region of about 1530 MeV seem to play a dominant role, but no consensus
is reached yet how large their importance actually is and whether the $N^*(1520)$ is 
the most important one or the $N^*(1535)$ or both (see also \cite{Shklyar:2004ba}). 

Clearly, the more structures we introduce, the more parameters we get. In turn, their
correlations deduced from the sum rules get more complicated. To keep things simple
we choose the following parameterization with two narrow peaks
\begin{equation}
  \label{eq:imrhad-tnp}
{\rm Im}R_{\rm HAD}(s,\rho_N) = 
2\pi (F_\pi^2 + c_F \rho_N) \, \delta(s-M_\omega^2-c_M \rho_N) 
+ 2\pi c_{rh} \rho_N \, \delta(s-M_{rh}^2)
\end{equation}
with
\begin{equation}
  \label{eq:rhmass}
M_{rh} = m_R - m_N
\end{equation}
and the resonance mass $m_R$. As pointed out above, a reasonable choice is
$m_R \approx 1530\,$MeV covering basically
the lowest $D_{13}$ and $S_{11}$ excitation, respectively.
Note that possible in-medium shifts of the resonance-hole peak, e.g.~by level repulsion,
are beyond the linear-density approximation used here. Of course, the parameterization
(\ref{eq:imrhad-tnp}) is less general than (\ref{eq:imrhad-owp}) concerning the aspect
that in (\ref{eq:imrhad-tnp}) the widths of the peaks are neglected. We would like to 
stress, however, that we want to keep the parameterizations as simple as possible.
In addition, we have included the present subsection to avoid the wrong
impression that mass and width would be the only things which one must consider
for a comprehensive discussion of an in-medium spectral information. 
\begin{figure}[htbp]
  \centering
    \includegraphics[keepaspectratio,width=0.49\textwidth]{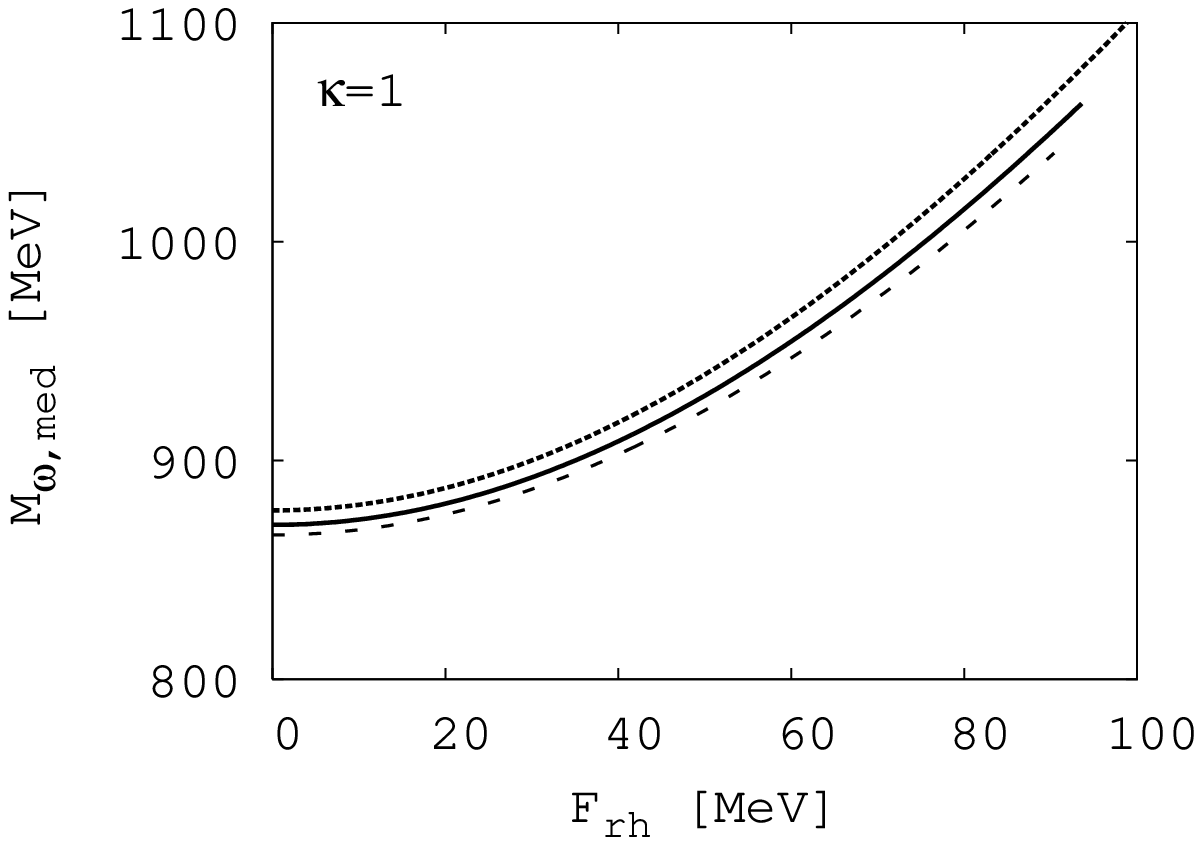}
    \includegraphics[keepaspectratio,width=0.49\textwidth]{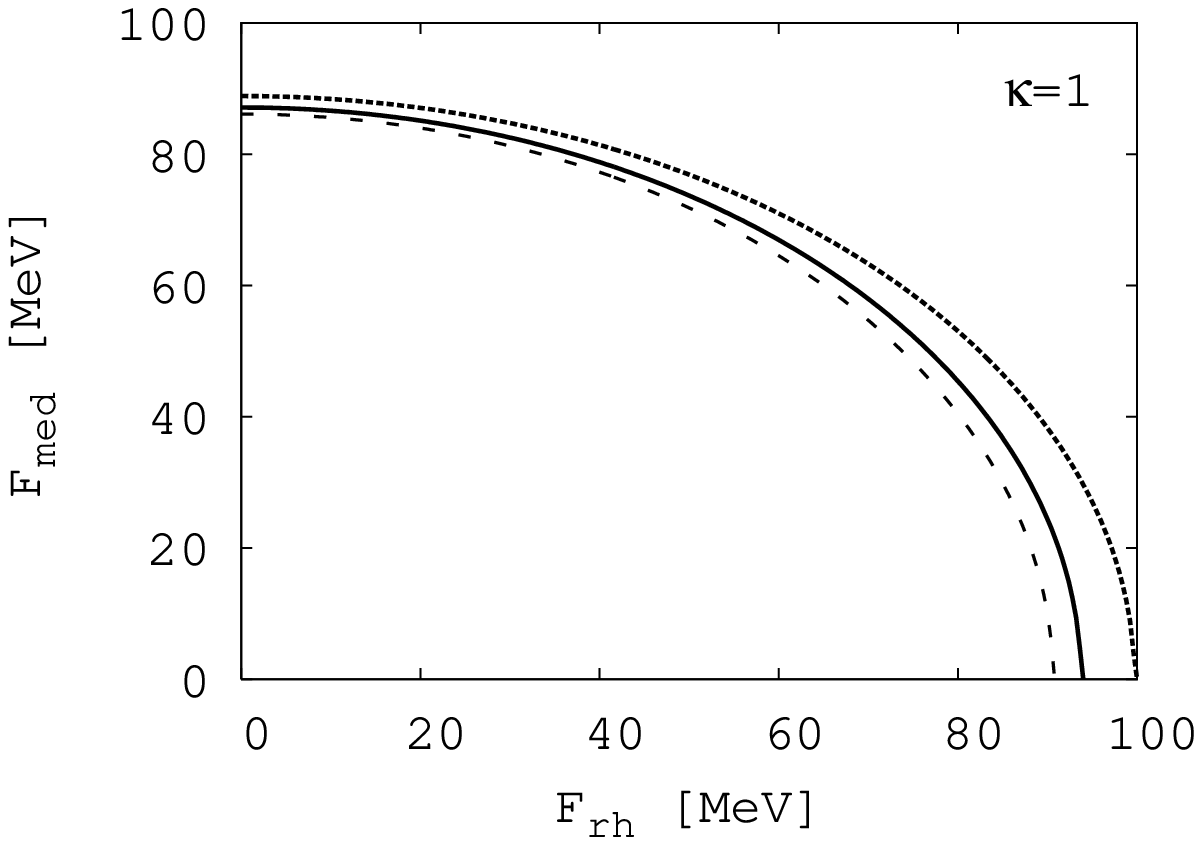}
  \caption{In-medium mass \eqref{eq:defmass} (left panel) and 
in-medium strength \eqref{eq:defstr} (right panel)
of the $\omega$-meson for normal nuclear matter density 
$\rho_N = \rho_0 \approx 0.17\,$fm$^{-3}$ as functions of the strength of the 
resonance-hole branch for different values
of the vacuum continuum threshold $s_0 = 1.33\,$GeV$^2$ (full lines), 
$1.20\,$GeV$^2$ (dotted), $1.46\,$GeV$^2$ (dashed). Note that the input is a structure with
two narrow 
peaks \eqref{eq:imrhad-tnp}. Also note that the lines in the left panel end 
at the respective value of $F_{rh}$ where in the right plot $F_{\rm med} =0$ is reached.
}
  \label{fig:twopeak}
\end{figure}

A two-peak structure has, of course, two strengths which is the strength for the genuine
$\omega$-branch as already defined in (\ref{eq:defstr}) and, in addition, the 
strength of the resonance-hole peak
\begin{equation}
  \label{eq:defstrrh}
F_{rh} = \sqrt{c_{rh} \rho_N}  \,.
\end{equation}

Again, we have three parameters.\footnote{Since we have neglected the 
widths of the states.} Given e.g.~the strength of the 
resonance-hole branch (\ref{eq:defstrrh}), the sum rules determine the in-medium mass
(\ref{eq:defmass}) and the in-medium strength (\ref{eq:defstr})
of the genuine $\omega$-branch. 
These correlations are depicted in figure \ref{fig:twopeak} for $\kappa =1$. 
Similar curves would emerge for other values of $\kappa$. All 
curves would be shifted downwards for increasing $\kappa$ (not explicitly shown).
In the right panel of figure \ref{fig:twopeak} we see how strength can be reshuffled from
one branch to the other. Collecting more strength in the resonance-hole branch, 
i.e.~at low invariant masses, however, results in a need for compensation. 
This is achieved by shifting the other branch upwards. We observe this effect in
the left panel of figure \ref{fig:twopeak}.

\section{Summary and outlook}
\label{sec:sum}

In the present work we have derived new sum rules for the $\omega$-meson in nuclear
matter, namely weighted finite energy sum rules. These sum rules are free of additional 
parameters (like Borel masses) which are not of primary interest. 
In addition, the derived sum rules are directly 
sensitive to the in-medium changes and do not mix vacuum and in-medium information.
We have performed sum rule analyses for different parameterizations of the spectral
information. In part, we have repeated the sum rule analysis of other groups
using our type of sum rules instead of theirs. In these cases, we could qualitatively
reproduce their results. Quantitative differences are due to the different sum rules 
used.
We have pointed out after equation (\ref{eq:kappalnc}) and 
after equations (\ref{eq:FESR}) why we prefer the use of weighted finite energy sum 
rules. 

It should have become clear from the present work that the sum rules alone cannot 
predict e.g.~an 
in-medium mass shift. Instead, the result of a QCD sum rule analysis depends
on the chosen input parameterization. Nonetheless, for a given parameterization
the sum rules can correlate the input parameters. Examples are presented in
figures \ref{fig:broadpeak} and \ref{fig:twopeak}. The most important qualitative
finding is that the sum rules prefer a shift of spectral strength to higher invariant
masses as compared to the vacuum case --- except for very large values of $\kappa$
which parameterizes the in-medium change of the four-quark condensate. If some spectral
strength is shifted to lower invariant masses --- by increasing the width of the
$\omega$-peak or by introducing a low-mass resonance-hole branch --- the mass
of the $\omega$-peak/branch moves further upwards. 

Apparently, an in-medium upwards shift of strength seems to be in conflict with the
experimental finding of a decreasing $\omega$-mass in the interior of 
a nucleus \cite{Trnka:2005ey}: 
On the theoretical side there are
several possibilities which could resolve this contradiction. First, the value for
$\kappa$ might be so large that the in-medium $\omega$-mass drops, 
cf.~figure~\ref{fig:narpeak}. Indeed, in \cite{Thomas:2005dc} the observed drop in
\cite{Trnka:2005ey} is taken as an indication that $\kappa$ seems to be large.
As pointed out above, $\kappa$ becomes 1 in the limit
of a large number $N_c$ of colors. If a very large value of $\kappa$ was the right 
explanation for an in-medium drop
of the $\omega$-mass, it would be interesting to understand why the large-$N_c$
approximation fails so badly. A second possible explanation would be that the
linear-density approximation used throughout this analysis breaks down already at
rather low densities. This approximation basically involves only vacuum physics
(e.g.~a selfenergy is given by $\rho_N T$ where $T$ is a {\em vacuum} scattering
amplitude, cf.~e.g.~the discussion in \cite{MuhlichShklyar:2005}). Beyond the 
linear-density approximation more complicated effects might play a dominant role.
Most spectacular is the proposed in-medium change of the underlying vacuum structure
as is formulated e.g.~in the scaling of hadron masses proposed by 
Brown and Rho \cite{Brown:1991kk} or in the hidden local symmetry approach
\cite{Harada:2003jx}. In such scenarios it is suggested that the vector meson mass drops.
Such effects might overwhelm the linear-density effect deduced here. It should be
stressed, however, that the effects discussed in the present work are not included
in the cited works either. A more complete treatment would be desirable. 
However, there is also a third possible explanation which is much less spectacular:
It might appear that only part of the spectral information has been identified in the 
experiment. Especially in a scenario with a sizable resonance-hole branch (discussed
in subsection \ref{sec:tnp}) it might turn out that only the low-mass resonance-hole
branch is seen whereas the rather high-lying (cf.~figure \ref{fig:twopeak}) 
genuine $\omega$-branch is missed. Note that the strength, i.e.~the visibility
of the genuine $\omega$-branch drops with increasing importance of the resonance-hole
branch, as shown in figure \ref{fig:twopeak}, right panel. 
It would be interesting, if the theoretical
approaches which try to describe the data of \cite{Trnka:2005ey} included at least
the resonance decay processes $N^* \to N \pi^0 \gamma$, since in \cite{Trnka:2005ey}
the $\omega$-meson is identified via its decay mode $\omega \to \pi^0 \gamma$.
Finally we note the possibility that one sees also the $\rho$ and not only the $\omega$
in the in-medium spectrum of $\pi \gamma$: In {\em vacuum} the decay channel 
$\rho \to \pi \gamma$ is very small as compared 
to $\omega \to \pi^0 \gamma$ \cite{pdg04}. However, this might change in the medium. 
Experimentally, this might be checked by looking in addition at the in-medium spectra
of charged final states,
i.e.~at $\rho^\pm \to \pi^\pm \gamma$ (which, of course, do not exist for the $\omega$).

Our analysis demonstrates that for a given
hadronic model it can be checked whether the model is consistent with the sum rules. 
If the hadronic model contains only a few undetermined or only roughly determined 
parameters they can be predicted or at least correlated. Note, however, that the
sum rules make statements about the current-current correlator and not directly about
the $\omega$-meson. As we have shown above, the most simplest connection, i.e.~strict 
vector meson dominance is excluded by the sum rules. We regard this as an important
point of the present sum rule analysis.

Concerning consistency checks for hadronic models it is also interesting to point out
that the first of our two used weighted finite energy sum rules, i.e.~(\ref{eq:wFESR1}),
is independent of the parameter $\kappa$, i.e.~independent of the four-quark condensate 
which is notoriously difficult to pin down. This sum rule provides a good check for
the consistency of a given hadronic model. We stress again that the obtained 
weighted finite energy sum rules are directly sensitive to the in-medium part of the
hadronic model and do not mix vacuum and in-medium information.

\acknowledgments S.L.~thanks V.~Metag for his questions which initiated this work.
He also acknowledges stimulating discussions with U.~Mosel, B.~K\"ampfer, M.~Kotulla,
R.~Thomas, D.~Trnka, W.~Weise and S.~Zschocke. 
S.L.~acknowledges the support of the European Community-Research Infrastructure
Activity under the FP6 ``Structuring the European Research Area'' programme
(HadronPhysics, contract number RII3-CT-2004-506078). The work of B.S.~is supported
by the DFG in the context of the European Graduate School ``Complex Systems of Hadrons 
and Nuclei''.

\appendix

\section{Landau damping contribution for 
  isovector and isoscalar currents}

One contribution to the in-medium expectation value of a current-current 
correlator comes from the Landau damping process. 
In the following this Landau damping contribution is calculated (in two ways) 
for the isoscalar and isovector current at rest with respect to the nucleons
which form the nuclear medium. Note that we 
work in the linear-density approximation. Therefore the in-medium interaction is 
approximated by a sum of single nucleon interactions with the current. 
The respective nucleon is at rest with respect to the medium.

We will start out with a model independent derivation. In addition, we discuss afterwards
a derivation using vector meson dominance and the universality of the coupling strengths
of the vector mesons. 

The currents appropriate for our discussion are the isovector current,
\begin{equation}
  \label{eq:currho}
j_\mu^\rho  := {1 \over 2} \, (\bar u \gamma_\mu u - \bar d \gamma_\mu d )  \,,
\end{equation}
and the isoscalar current,
\begin{equation}
  \label{eq:curom}
j_\mu^\omega  := {1 \over 2} \, (\bar u \gamma_\mu u + \bar d \gamma_\mu d ) \,.
\end{equation}
The currents chosen in such way yield in vacuum the same operator product expansion.
Next we have to figure out how an electromagnetic current is decomposed into the
previous currents: 
The electromagnetic current is (neglecting strange and more heavier quarks)
\begin{equation}
  \label{eq:emcur}
j_\mu^{\rm el} = Q_u \, \bar u \gamma_\mu u + Q_d \, \bar d \gamma_\mu d 
= j_\mu^\rho + \frac13 \, j_\mu^\omega
\end{equation}
Following \cite{Hatsuda:1995dy} the nucleon matrix element of the isospin-1 part of the
electromagnetic current has two form factors:
\begin{equation}
  \label{eq:hatsapp}
\langle N(\vec k_1) \vert j_\mu^{\rho} \vert N(\vec k_2) \rangle
= \bar u(\vec k_1) \, {\tau^3 \over 2} \,
[F^{I=1}_1(q) \gamma_\mu + F^{I=1}_2(q) i \sigma_{\mu\nu} q^\nu ] \, u(\vec k_2) \,.
\end{equation}
The corresponding relation for the isospin-0 part is
\begin{equation}
  \label{eq:hatsapp0}
\langle N(\vec k_1) \vert {1 \over 3} \, j_\mu^{\omega} \vert N(\vec k_2) \rangle
= \bar u(\vec k_1) \, {1 \over 2} 
[F^{I=0}_1(q) \gamma_\mu + F^{I=0}_2(q) i \sigma_{\mu\nu} q^\nu ] \, u(\vec k_2) \,.
\end{equation}
For the current at rest with respect to the nuclear medium, i.e.~for $\vec q = 0$,
the energy for the Landau damping process is fixed to 
$q_0=0$ \cite{Hatsuda:1995dy}. In this case, the contribution from the tensor
part vanishes. On the other hand, the $F_1$ form factors become unity. This yields
a model independent result for the Landau damping contribution to the isoscalar
sum rule: 
\begin{equation}
  \label{eq:landaufinalom}
- {9 \over 4 M^2} \, {\rho_N \over m_N} \,.
\end{equation}
It is included in (\ref{eq:sumrule}) as the second term on the right hand side.
For more details we refer to \cite{Hatsuda:1995dy}. Concerning the isovector case
we get in the same way:
\begin{equation}
  \label{eq:landaufinalrho}
- {1 \over 4 M^2} \, {\rho_N \over m_N}  \,.
\end{equation}
The relative factor 9 between
isovector and isoscalar has also
been used in \cite{klingl2,Dutt-Mazumder:2000ys,Zschocke:2002mp}.

A second derivation of the same results uses vector meson dominance: We start with
Lagrangians \cite{klingl2} for $\rho$-nucleon interaction,
\begin{equation}
  \label{eq:rhonlagr}
{\cal L}_{\rho N} = g_{\rho N} \bar N \tau^a \gamma_\mu N \rho_a^\mu +
{g_{\rho N} \kappa_\rho \over 4 m_N} 
\bar N \tau^a \sigma_{\mu\nu} N \partial^\mu \rho^\nu_a  \,,
\end{equation}
and $\omega$-nucleon interaction, respectively,
\begin{equation}
  \label{eq:omnlagr}
{\cal L}_{\omega N} = g_{\omega N} \bar N \gamma_\mu N \omega^\mu +
{g_{\omega N} \kappa_\omega \over 4 m_N} 
\bar N \sigma_{\mu\nu} N \partial^\mu \omega^\nu  \,.
\end{equation}
Assuming a universal coupling we get \cite{klingl2}
\begin{equation}
  \label{eq:coupl}
g_\omega = 3 g_\rho \,, \quad g_{\omega N} = g_\omega \,, \quad 
g_{\rho N} = g_\rho \,.
\end{equation}
Using vector meson dominance the electromagnetic current is given by
\begin{equation}
  \label{eq:emvmd}
j_\mu^{\rm el} = 
-{m_V^2 \over g_\rho} \, \rho_\mu - {m_V^2 \over g_\omega} \, \omega_\mu
= -{m_V^2 \over g_\rho} \, \left( \rho_\mu + {1 \over 3} \omega_\mu \right) \,.
\end{equation}
This result has to be compared to (\ref{eq:emcur}) yielding
\begin{equation}
  \label{eq:veccur}
j_\mu^\rho = -{m_V^2 \over g_\rho} \, \rho_\mu \,, \qquad
j_\mu^\omega = -{m_V^2 \over g_\rho} \, \omega_\mu  \,.
\end{equation}
For our case $q=0$ (cf.~discussion above) the tensor interactions vanish and only the 
vector interactions contribute. We get
\begin{equation}
  \label{eq:nuclrho}
\langle N(\vec k_1) \vert j_\mu^\rho \vert N(\vec k_2) \rangle \sim 
{1 \over g_\rho} \, \langle N(\vec k_1) \vert \rho_\mu \vert N(\vec k_2) \rangle
\sim {g_{\rho N} \over g_\rho} = 1 \,,
\end{equation}
but
\begin{equation}
  \label{eq:nuclom}
\langle N(\vec k_1) \vert j_\mu^\omega \vert N(\vec k_2) \rangle \sim 
{1 \over g_\rho} \, \langle N(\vec k_1) \vert \omega_\mu \vert N(\vec k_2) \rangle
\sim {g_{\omega N} \over g_\rho} = 3 \,.
\end{equation}
These results are in agreement with the ones derived above, (\ref{eq:hatsapp}) and
(\ref{eq:hatsapp0}).

\bibliography{literature,priv}
\bibliographystyle{apsrev}

\end{document}